\documentclass[11pt]{article}
\begin{document}
\title{Universal Statistics of Inviscid Burgers Turbulence \\
in Arbitrary Dimensions}
\author{Ali Naji$^1$ and Shahin Rouhani$^{1,2}$\\
\\
$^1$Department of Physics, Sharif University of Technology,\\
Tehran, P.O.Box: 11365-9161, Iran\\
$^2$Institute for Studies in Theoretical Physics and Mathematics,\\
Tehran, P.O.Box: 19395-5531, Iran\\
e-mail: naaji$@$physics.sharif.ac.ir}
\date{\today}
\maketitle

%%%%%%%%%%%%%%%%%%%%%%%%%%%%%%%%%%%%%%%%%%%%%%%%%%%%%%%%%%%%%%%%%%%%%%%%%%%%%
\begin{abstract}
We investigate the non-perturbative results of multi-dimensional forced
Burgers equation coupled to the continuity equation. 
In the inviscid limit, we derive the exact 
exponents of two-point density correlation functions  
in the universal region in arbitrary dimensions.
We then find the universal generating function and the 
tails of the probability density function (PDF) 
for the longitudinal velocity difference.
Our results exhibit that in the inviscid limit, density fluctuations affect 
the master equation of the generating function  
in such a way that we can get a positive PDF with the well-known exponential
tail. The exponent of the algebraic tail is derived to be $-5/2$ in any dimension. 
Finally we observe that various forcing spectrums do not alter the power law 
behaviour of the algebraic tail in these dimensions, due to a 
relation between forcing correlator exponent and the exponent of the two-point density
correlation function.\\
PACS numbers 47.27.Gs and 47.40.Ki 
\end{abstract}

%\hspace{7.30mm}PACS numbers 47.27.Gs and 47.40.Ki 
%\newpage
%%%%%%%%%%%%%%%%%%%%%%%%%%%%%%%%%%%%%%%%%%%%%%%%%%%%%%%%%%%%%%%%%%%%%%%%%%%%%
The interest in solving the randomly 
driven Burgers equation with a large scale driving force is motivated 
by the hope that it can provide us with the first solvable model 
of turbulence. 
Consequently tremendous activity emerged on the 
non-perturbative understanding of Burgers turbulence 
\cite{Pol,Bol1,Bol2,Gu,BMP,BM,Kh,E,Kr,Go1,BD,MRRT}. 
As an original attempt, Polyakov offered a field theoretical method
to derive the probability distribution or density function (PDF)
of velocity difference in the problem of randomly
driven Burgers equation in one dimension \cite{Pol}.
The problem of computation of correlation functions 
in the inertial range is reduced to the solution of a closed partial 
differential equation \cite{Pol,Bol1}. However, Polyakov's approach
was based on the conjecture of the existence of the operator product
expansion (OPE). 
This method was then extended to the forced Burgers equation coupled to the 
continuity equation and some results were derived for the 
behaviour of the probability density function tails 
and for the value of intermittency and density-density correlators     
exponents \cite{Bol2,MRRT}.
On the other hand, some extensive numerical simulations show that 
the predictions of these theoretical works coincide with the numerical
simulations, within a good approximation \cite{Go1,Ch}.

It was known \cite{Pol,Bol1,Bol2,Gu,BMP,BM,Kh,E,Kr,Go1,MRRT,Ch}
that the PDF in Burgers turbulence behaves physically different in two 
different intervals of velocity difference $u$. When $|u|>U_{rms}$ 
the PDF behaves as:
\begin{eqnarray}
P(u,r)=rG(\frac{u}{U_{rms}}) \nonumber
\end{eqnarray}
which depends on the single-point property $U_{rms}$ (root-mean-square 
of velocity fluctuations) and therefore is not a universal function. 
While in the interval $u \ll U_{rms}$ and $r \ll L$ 
(where $L$ is the energy input scale), the PDF can be represented 
in the universal scaling form:
\begin{eqnarray}
P(u,r)=\frac{1}{r^{\delta}} F(\frac{u}{r^{\delta}}) \nonumber
\end{eqnarray}
where $\int^{+\infty}_{-\infty}F(x)dx=1$ and the exponent $\delta$ is related to
the random force spectrum \cite{Ch}. Results for one-dimensional 
case indicates that the function $F(x)$ behaves like $\exp[-c(\delta)x^3]$
when $x\rightarrow +\infty$, where the coefficient $c(\delta)$ can be evaluated from the
theory \cite{E,Ch}.
In this region, various algebraic behaviours were predicted for 
the PDF in the limit $x\rightarrow -\infty$ and determination of 
the asymptotic behaviour of the PDF is at the moment controversial. 
Several different proposals have been made, each leads to an asymptotic
experssion of the forms $\sim |u|^{-\alpha}$ but with a variety of values for
the exponent $\alpha$ includes $2$ \cite{BM,Ch}, the range $5/2$ to $3$
\cite{Pol,Bol1,MRRT}, $3$ \cite{Go1} and $7/2$ \cite{Kh,E,Kr}.   
Numerical simulations performed in \cite{Ch} shows that the power law 
exponent of the PDF depends on the forcing spectrum. 
The investigations of forced Burgers turbulence models
have also given furthur understanding of intermittency in the velocity
structure functions. It is believed that intermittency in these systems                          
is a consequence of the algebraic tail of PDF implying the scaling exponent
of velocity structure functions, $\xi_n$, is saturated to the value $\xi_n=1$
for $n>n_c$, where the value of critical moment number $n_c$ depends on the 
forcing function spectrum \cite{Ch}. This simply means that the moments 
$S_{n>n_c}$ are dominated by the non-universal region and thus vary with the 
sigle-point property $U_{rms}$ induced by the large scale random force. 
At the same time, the moments with $n<n_c$ are universal \cite{Pol,Ch}. 
  
The first attempt on $d$-dimensional Navier-Stokes turbulence was made
in \cite{FF} as a renormalized perturbation expansion with $1/d$ as a
small parameter. This attempt failed because $1/d$ factors appearing       
in the expansion are cancelled by the $O(d)$ multipliers resulting from
the summation over the $d$ components of velocity field.
The problem of $d$-dimensional turbulence was recently revisited \cite{Yc} 
using Polyakov's non-perturbative approach. It was shown 
that in the limit $d \rightarrow \infty$ 
Kolmogorov scaling is an exact solution of the incompressible Navier-Stokes 
equations and the pressure contributions to the equations are responsible
for a clear distinction between the Navier-Stokes and Burgers dynamics. 
In an other work \cite{Bol2}, the asymptotics of the PDF of velocity gradient      %Bol2 
were found in $d$-dimensional randomly driven Burgers                              %ilucidation
equation in a compressible fluid. 

In the present paper we investigate
the isotropic forced Burgers equation coupled to the continuity equation in 
arbitrary dimensions. 
In the inviscid limit, we find a complete closed 
equation for the generating function of velocity difference.
We solve this equation for the longitudinal component of velocity 
difference in the universal region and show that the main requirments 
on the PDF fix the exponent of two-point density correlation function. This results
in a dimension dependent exponent for density fluctuations. This then means that the 
equation governing the PDF is independent of dimensions.  
In this region, our results predict the exponential decaying of the right 
tail of PDF of velocity difference which is also obtained in several other 
works \cite{Pol,Bol1,Bol2,Gu,BM,E,Kr,Go1,MRRT}. Thus the power law decay 
of the left tail is independent of the dimensionality and the forcing 
spectrum with the exponent $-5/2$ in any dimension.
While the previous results for one-dimensional forced Burgers turbulence
exhibit a force spectrum dependent tail \cite{Pol,Bol1,Ch} or
a different power law decaying \cite{BM,Kh,E,Kr,Go1}.  
Finally, we discuss Polyakov's approach \cite{Pol} for dealing with the problem in the
limit of infinitesimal viscosity. For a special forcing spectrum this method 
would not give anything more than the inviscid calculations.                            
Our work extends a previous result \cite{MRRT} on two and three-dimensional                %*******
Burgers turbulence to higher dimensions.                     
\vspace{0.3in}

Let us start with the Burgers and continuity equations:

\begin{equation}
{\partial}_t{\bf u} + ({\bf u}\cdot\nabla){\bf u}=\nu\nabla^2{\bf u}
+{\bf f}({\bf x},t)
\end{equation}
\begin{equation}
{\partial}_t{\rho}+\nabla \cdot(\rho{\bf u})=0
\end{equation}
for the irrotational velocity field ${\bf u}({\bf x},t)$, viscosity
$\nu$ and density $\rho$ in $d$ dimensions. The force ${\bf f}({\bf x},t)$
is the external stirring force, which injects energy into the system
on a length scale $L$. More specifically, one can take, for instance
a Gaussian distributed random force, which is identified by its
two moments:

\begin{eqnarray}
{\langle f_\mu({\bf x}, t)\rangle}=0 \nonumber
\end{eqnarray}
\begin{equation}
\langle f_\mu ({\bf x},t) f_\nu ({\bf x'},t') \rangle= 
\delta(t-t')k_{\mu\nu}({\bf x-x' })
\end{equation}
where $\mu,\nu=1,2,\ldots,d$ and the correlation function
$k_{\mu\nu}(r)$ is concentrated at some large scale $L$.
The problem is to understand the statistical properties of the velocity
and density fields which are the solutions of equations (1) and (2).
Following Polyakov, we consider the following two-point
generating functional:

\begin{equation}
F_2(\lambda_1,\lambda_2,{\bf x}_1,{\bf x}_2,t)=
\langle \rho({\bf x}_1,t)\rho({\bf x}_2,t)\exp
(\lambda_1\cdot{\bf u}({\bf x}_1,t)+\lambda_2\cdot{\bf u}
({\bf x}_2,t))\rangle
\end{equation}
where the symbol $\langle\ldots\rangle$ means an average over
various realization of the random force.
Now one can show that ${F}_{2}$ satisfies
the following equation:

\begin{eqnarray}
\partial_t F_2+
\sum_{i=1,2; \mu=1,\ldots,d} \frac{\partial}
{\partial \lambda_{\mu,i}} \partial_{\mu,i}F_2
-\nonumber\\
\sum_{i,j=1,2; \mu,\nu=1,\ldots,d} \lambda_{\mu,i}\lambda_{\nu,j}
k_{\mu\nu}({\bf x}_i -{\bf x}_j)F_2=D_2
\end{eqnarray}
where the first two terms on the left-hand side of equation (5)
come from the terms on the left-hand side of equations (1) and (2),
and the third is the contribution of forcing
term, in which we have used Furutsu-Novikov-Donsker formula \cite{Woy}.
Also, $D_2$-term is the contribution of dissipation.
$D_2$ is the anomaly term and has the following form:

\begin{equation}
D_2=\langle\nu\rho({\bf x}_1)\rho({\bf x}_2)
[\lambda_1\cdot\nabla^2{\bf u}({\bf x}_1)+
\lambda_ 2\cdot\nabla^2{\bf u}({\bf x}_2)]
\exp(\lambda_1\cdot{\bf u}({\bf x}_1)
+\lambda_2\cdot{\bf u}({\bf x}_2))\rangle
\end{equation}
It is noted that although the advection contribution are
accurately accounted for in this equation, it is not closed due to the
dissipation term. 
In what follows, we consider the inviscid ($\nu = 0$)
Burgers equation to avoid the anomaly problem and
leave Polyakov's approach for dealing with the problem in the
limit $\nu\rightarrow 0$ to the last part of the present paper.
Now we change the variables as ${\bf x}_\pm
={\bf x}_1 \pm {\bf x}_2$, $\lambda_+=\lambda_1+\lambda_2$ and
$\lambda_-=\frac{\lambda_1-\lambda_2}{2}$.
By the Galilean invariance and the spatial homogeneity assumptions the
variables ${\bf x}_+$ and $\lambda_+$ can be set to zero.
It is found from the equation (5) that in the {\it stationary state}
we have:

\begin{equation}
{\sum_{\mu=1,\ldots,d}}{\frac {\partial}{{\partial \lambda}_\mu}}
{\frac {\partial}{{\partial x}_{\mu}}}{F_2}
+2{\sum_{\mu,\nu=1,\ldots,d}}\lambda_{\mu}\lambda_{\nu}
[k_{\mu\nu}({\bf x})-k_{\mu\nu}(0)]
{F_2}=0
\end{equation}
where we have used ${\bf x}$ and $\lambda$ instead of ${\bf x}_-$ and
$\lambda_-$ for simplification. Also, because of isotropy
$F_{2}$ can depend only on the absolute values of vectors
${\bf x}$ and $\lambda$ and the angle $\theta$ between them
as $F_{2}={F_2}(r,\lambda,s)$ where
$r=|{\bf x}|$, $\lambda=|{\lambda}|$             
and $s={\cos}\theta=(\sum{x_\mu}{\lambda_\mu})/{r}{\lambda}$.
We suppose the stirring correlation function 
has the following form:

\begin{equation}
k_{\mu\nu}({\bf x})=
k_0\delta_{\mu\nu}-\frac{k_1}{2L^2}r^2(\delta_{\mu\nu}
+2\frac{x_\mu x_\nu}{r^2})
\end{equation}
with $k_1$ and $L=1$.
Now, using spherical coordinates $(r,\lambda,s)$ it can be shown
from equations (7) and (8) that $F_{2}$ satisfies the following
closed equation for homogeneous and isotropic turbulence:       

\begin{eqnarray}
[{s}{\partial _r}{\partial _{\lambda}}-{\frac {{s}(1-{s}^2)}{r{\lambda}} 
{\partial^2_{s}}+{\frac {(d-2+{s}^2)}{r{\lambda}} }{\partial _{s}}
+{\frac {(1-{s}^2)}{\lambda} }{\partial_r}{\partial_{s}}}\nonumber\\
+{\frac {(1-{s}^2)}{r} }{\partial_{\lambda}}{\partial_{s}}
-r^2{\lambda}^2(1+2{s}^2)]{F_2}=0
\end{eqnarray}
The one-dimensional case of equation (7) is easily recovered by setting $s=1$.
We wish to consider the longitudinal velocity component statistics
which results from equation (9) by taking the limit $s\rightarrow 1$.
Assuming ${{F_2}}$ be a regular function near $s=1$ we can safely drop
the terms multiply by $(1-s^2)$ in equation (9).
We propose the universal scale invariant
solution of equation (9) in the following form:

\begin{eqnarray}
{F}_{2}{(r,{\lambda},s)}=g(r){F}
{({\lambda}{r^{\delta}},s)}\nonumber
\end{eqnarray}
\begin{equation}
g(r)={r^{-\alpha_d}}
\end{equation}
In equation (10), $g(r)$ is the conditional two-point
correlation function of density field conditioned on a fixed value
of velocity differnce. We assume that $g(r)$ depends only on $r$
and its dependence on the velocity interval appears in the exponent
$\alpha_d$, we shall discuss this further in the next section.
For a general stirring correlation function
$k_{\mu\nu}{\sim}(1-r^\zeta)$, the exponent $\delta$ is found by
substituting the proposed form of generating function as
$\delta=\frac{\zeta+1}{3}$ which in our case ($\zeta=2$) is $\delta=1$. 
Indeed, we assume that the two-point density correlation
function exists, and therefore it is necessary to find 
${F}{({\lambda}{r^{\delta}},s)}$ such that it tends to
a constant in the limit of $\lambda\rightarrow 0$.
Also, It is straight forward to show that the mean value of 
velocity difference is zero as expected from the homogeneity 
and isotropy constraints.
Now let us consider the longitudinal velocity component statistics
suggested by equation (9) in the limit $s\rightarrow 1$.
In this limit, we assume that the generating function of velocity difference
$F({\lambda}r,s)$ has the following form:

\begin{eqnarray}
{F}({\lambda}r,s)=F({\lambda}rs)
\end{eqnarray}
this form ensures the factorizing property of the angular part of
structure functions as $S_n(r,s)\sim s^n S_n(r)$ when $n<1$. The
factorization of the angular part of velocity structure functions
in the limit $s\rightarrow 1$ has also been known for the Navier-Stokes 
turbulence \cite{YY}. Plugging the ansatz (10) and (11) into equation (9) 
and rewriting it for the variable $z=\lambda rs$ gives the following
equation for the generating function of longitudinal velocity difference:

\begin{eqnarray} 
zF''(z)+(d-\alpha_d)F'(z)-3z^2F(z)=0
\end{eqnarray}
It is interesting that equation (12) is similar to the equation
first derived by Polyakov \cite{Pol} for the problem of 
one dimensional Burgers. In that work the effect of the viscous 
term is found by taking the limit of infinitesimal viscosity applying the
self-consistent conjectures of operator product expansion.
The anomaly terms which arise from the viscous term modify the master 
equation of the generating function in such a way that a positive, finite
and normalizable PDF can be found.
Comparing equation (12) with Polyakov's
result shows that the anomaly term in Polyakov's approach is replaced with
the exponent of two-point density correlation function such that a simple 
change of the parameters maps equation (12) to one derived
by Polyakov \cite{Pol}.
Therefore,
as mentioned in \cite{MRRT},
the presented approach for inviscid forced Burgers turbulence
shows that considering the density fluctuations 
coupled to the velocity field, 
alters the governing equations in such a way that we can obtain positive 
PDF even in the inviscid limit and with no need for the viscosity anomaly. 
It is easy to show using equation (12),
as discussed in \cite{Pol}, that the main requirments on the PDF forces us 
to take the two point density correlator exponents in arbitrary dimensions as:

\begin{eqnarray}
\alpha_d=d+\frac{1}{2}
\end{eqnarray}
which yields $F(z)=\exp[\frac{2}{\sqrt{3}}(z^{3/2})]$.
This result consequently gives the right tail of the PDF of 
longitudinal velocity differrence as $\frac{1}{r}\exp[-c(\frac{u}{r})^3]$
(for $\frac{u}{r}\rightarrow +\infty$) where $c=\frac{1}{9}$. 
This tail coincides with the result of several other approaches
\cite{Pol,Bol1,Bol2,Gu,BM,E,Kr,Go1}.
The left tail of the PDF is obtained in the limit of $\frac{u}{r}\rightarrow -\infty$
as $|u|^{-(\alpha_d-d+2)}$ or, by equation (13), $|u|^{-5/2}$.
As a result, density fluctuations which couple to the velocity field
appear themselves in the exponent of left tail of PDF 
and lead to a dimensionality independent decay law.
\vspace{0.3in}

Now we wish to focus our attention on the dependnce of these results on 
the stirring force spectrum. We consider a general large-scale stirring 
force correlator as follows:

\begin{eqnarray}
k_{\mu\nu}({\bf x})=k_0\delta_{\mu\nu}-\frac{1}{2}r^{\zeta}(\delta_{\mu\nu}
+2\frac{x_{\mu}x_{\nu}}{r^2})   
\end{eqnarray}
With this choice the velocity field remains irrotational also the 
coefficients in expression (14) are taken such it simply reduces to 
equation (8) when $\zeta=2$ .
The last term in the equation (5) can be written in the form
$-ar^{\zeta}\lambda^2 F_2$ for a general form of the stirring force.  
The coefficient $a$ depends on the parameters in the expression 
of the stirring correlator. For the choice as
in equation (14) $a$ is $(1+2s^2)$. 
However the exact form of $a$ does not affect the main features of the
problem such as algebraic tail of the PDF. 
Taking the universal solution (10) for the new master equation gives the
following eqation for the generating function $F(\lambda r^{\delta}s)=F(z)$
in the limit $s\rightarrow 1$:

\begin{eqnarray}
\delta z F''(z)+(d-\alpha_d+\delta-1)F'(z)-3z^2F(z)=0
\end{eqnarray}
where $\delta=\frac{1+\zeta}{3}$. The PDF of longitudinal velocity difference
is the inverse Laplace transform of $F(z)$. The positivity, finitness and
normalizability requirements on the PDF will fix the exponent of two-point density 
correlator as:

\begin{eqnarray}
\alpha_d=d+\frac{1}{2}(\zeta-1)
\end{eqnarray}
also we obtain $F(z)=\exp[\frac{2}{\sqrt{3\delta}}(z^{3/2})]$.
The right tail of the PDF can be readily deduced as $\exp[{-cx^3}]$
when $x=\frac{u}{r^{\delta}}\rightarrow +\infty$ and $c=-\frac{\delta}{9}$. 
On the other hand, the limit $x\rightarrow -\infty$ 
gives the left tail in the following form:

\begin{eqnarray}
P(u,r)\sim |u|^{-(\alpha_d-d+\delta+1)/{\delta}}
\end{eqnarray}
Substituting the value of $\alpha_d$ from equation (16) and
$\delta=\frac{\zeta+1}{3}$, we obtain the algebraic tail as:

\begin{eqnarray}
P(u,r)\sim  r^{3\delta/2}|u|^{-5/2}
\end{eqnarray}
We observe that the density fluctuations affect the algebraic tail of the 
PDF and modify its decaying exponent such that it becomes independent
of the dimension and the forcing spectrum.
\vspace{0.3in}

As mentioned previously, in the limit $\nu \rightarrow 0$, Polyakov 
formulated a method for analyzing the inertial range statistics based on the
conjecture of the existence of OPE. Extending the assumptions of OPE to take
into account the anomaly term in the case of arbitrary dimensions for Burgers 
equation coupled to the continuity equation, it was shown that 
$D_2$-term has the following structure \cite{Bol2}:

\begin{equation}
D_2=aF_2
\end{equation}
where $a$ is generally a function of $\lambda_1$ and $\lambda_2$.
Therefore keeping the viscosity infinitesimal but nonzero produces
a finite effect and a new term on the right hand side of equation (9) as:
\begin{eqnarray}
[{s}{\partial _r}{\partial _{\lambda}}-{\frac {{s}(1-{s}^2)}{r{\lambda}} 
{\partial^2_{s}}+{\frac{(d-2+{s}^2)}{r{\lambda}}}{\partial _{s}}
+{\frac{(1-{s}^2)}{\lambda}}{\partial_r}{\partial_{s}}}\nonumber\\
+{\frac{(1-{s}^2)}{r}}{\partial_{\lambda}}{\partial_{s}}
-r^2{\lambda}^2(1+2{s}^2)]{F_2}=a(\lambda){F_2}
\end{eqnarray}
The $\lambda$ dependence of $a(\lambda)$ anomaly must be chosen to conform
the scaling and can be changed depending on the properties of the force 
correlation function. For a general correlation function $k_{\mu \nu}{\sim}
{(1-{r}^{\zeta})}$ the $\lambda$ dependence of $a$ is fixed as 
$a_0{\lambda}^{\sigma}$ where $\sigma=\frac{2-\zeta}{1+\zeta}$ and $a_0$ 
is a constant. It is evident that in the case of $\zeta=2$, $a(\lambda)$ is 
independent of $\lambda$. In this case, one can easily show from the master 
equation that the parameter $a_0$ depends linearly on the mean value of 
velocity difference field of flow and therefore vanishes in the homogeneous
isotropic turbulence. However for different types of correlations of the 
stirring force, e.g.$ k_{\mu\nu}{\sim}{(1-{r}^{\zeta})}$ with $\zeta\neq 2$, we have to
assume non-zero $a_{0}$ \cite{Bol1}.
Thus, for the stirring correlation as the type $1-r^2$,
the results of Polyakov's formalism (in the limit of infinitesimal 
viscosity) will not be different from the inviscid results of homogeneous 
isotropic turbulence.

To our knowledge, there is a little informations about the statistics of the
density field. There exists one simulation \cite{Go2}
to find the PDF tails of density field for one dimensional decaying Burgers 
turbulence in the zero viscosity limit, which exhibits a power law tail
for the density PDF in the high density regime. 
In one dimension, Boldyrev \cite{Bol2} has reported a simulation
in which the exponent of two-point density correlation function
is predicted to be $\sim 2$. 
\vspace{0.3in}

We would like to thank J. Davoudi,  S. Moghimi-Araghi
and M.R. Rahimi Tabar for useful discussions.
%%%%%%%%%%%%%%%%%%%%%%%%%%%%%%%%%%%%%%%%%%%%%%%%%%%%%%%%%%%%%%%%%%%%%%%%%%%%%

%%%%%%%%%%%%%%%%%%%%%%%%%%%%%%%%%%%%%%%%%%%%%%%%%%%%%%%%%%%%%%%%%%%%%%%%%%%%%
\end{document}